\documentclass[12pt,preprint]{aastex}
\slugcomment{Accepted by the Astrophysical Journal}
\shorttitle{A Cluster of Hidden Outflows in the OMC1S Region}
\shortauthors{Zapata et al.}
\begin{document}

\title{Silicon Monoxide Observations Reveal a Cluster of Hidden 
Compact Outflows in the OMC1 South Region}

\author{Luis A. Zapata\altaffilmark{1,2}, Paul T. P. Ho\altaffilmark{2,3,4}, 
Luis F. Rodr\'\i guez\altaffilmark{1}, C. R. O'Dell\altaffilmark{5}, 
Qizhou Zhang\altaffilmark{2}, and August Muench\altaffilmark{2} } 
\altaffiltext{1}{Centro de Radioastronom\'\i a y Astrof\'\i sica,
 UNAM, Apdo. Postal 3-72 (Xangari), 58089 Morelia, Michoac\'an, M\'exico}
\altaffiltext{2}{Harvard-Smithsonian Center for Astrophysics, 60 Garden Street,
Cambridge, MA 02138, USA}
\altaffiltext{3}{Academia Sinica Institute of Astronomy and Astrophysics,
Taipei, Taiwan}
\altaffiltext{4}{Department of Physics, National Taiwan University, Taipei, Taiwan}
\altaffiltext{5}{Department of Physics and Astronomy, Vanderbilt University, 
Box 1807-B, Nashville, TN 37235}

\email{lzapata@astrosmo.unam.mx}
 
\begin{abstract}
We present high angular resolution ($2\rlap.{''}8 \times 1\rlap.{''}7$) 
SiO J=5$\rightarrow$4; $v=0$ line observations of the OMC1S 
region in the Orion Nebula made using the Submillimeter Array (SMA). We 
detect for the first time a cluster of four compact bipolar and monopolar 
outflows that show high, moderate and low velocity gas and appear 
to be energized by millimeter and infrared sources associated with 
this region. The SiO molecular outflows are compact 
($<$ 3500 AU), and in most of the cases, 
they are located very close to their exciting sources. 
We thus propose that the SiO thermal emission is tracing the
youngest and most highly excited parts of the outflows which cannot be detected 
by other molecules. Moreover, since the ambient cloud
is weak in the SiO line emission, these observations can reveal flows 
that in other molecular transitions
will be confused with the ambient velocity cloud emission. 
Analysis of their positional-velocity diagrams show that some components 
of these outflows may be driven by wide-angle winds very close to the 
exciting object.
Finally, we find that some of these SiO outflows seem to be the 
base of powerful Herbig-Haro 
jets and large-scale molecular flows that emanate from a 
few arcseconds around this zone. 
In particular, we find a strongly excited SiO bipolar outflow 
with a P.A. of $\sim$ 100$^{\circ}$, that is likely 
energized by the luminous ($\sim$ 3 $\times$ 10$^3$ L$_{\odot}$) infrared protostar
 "B" and could be the base of the remarkable object HH269.  
\end{abstract}  


\section{Introduction}

The OMC1S region (Orion Molecular Cloud 1 South) located about 1$'$ south-east of 
the Trapezium region, is one of the three most important zones of star formation 
(BN-KL, The Trapezium and OMC1S) in the Orion Nebula.
This region indeed seems to be the youngest and most active zone of star 
formation in the nebula, showing extremely well collimated molecular outflows, 
and at least a half-dozen extended and powerful HH flows emanating from it 
(Schmid-Burgk et al. 1990, Bally, O'Dell \& McCaughrean 2000, 
Rosado et al. 2001, O'Dell \& Doi 2003, Zapata et al. 2005). 

The OMC1S region was first identified as a star formation region through measurements 
of NH$_{3}$ (J,K)= (1,1) and (2,2) lines (Ziurys et al. 1981) and through 400 $\mu$m 
dust continuum images (Keene, Hildebrand \& Whitcomb 1982).
Later, NH$_3$ line observations showed very high gas kinetic temperature ($\geq$ 100 K),
possibly due to the presence of strong outflows associated with the region 
(Batrla et al. 1983; Mauersberger et al. 1986; Ziurys, Wilson \& Mauersberger 1990, 
Wiseman \& Ho 1998). 
Many molecular lines have been detected toward the OMC1 South region: 
SiO, SO$_2$, H$^{13}$CO$^+$, CH$_3$OH, CH$_3$CN, 
HC$_3$N, $^{34}$CS, H$_2$O (as maser emission), $^{12}$CO and $^{18}$CO (Batrla et al. 1983; 
Mundy et al. 1986; Ziurys, Wilson \& Mauersberger 1990; Schmid-Burgk et al. 1990; 
McMullin et al. 1993, Gaume et al. 1998 and Rodr\'\i guez-Franco et al. 1999a). 
However, as all of the species were detected with single-dish observations of moderate 
angular resolution, their relation to the embedded sources was not clear.

In particular, the thermal emission of the SiO line has provided 
evidence for the strong outflow activity associated with this region. 
Ziurys, Wilson \& Mauersberger (1990) found a low velocity (10 km s$^{-1}$)
and extended bipolar SiO(J=5-4; v=0) outflow with a length of $\sim$ 30${''}$ 
in the North(blueshifted)-South(redshifted) orientation, that is centered 
at the position R.A.[J2000]=$05^h~35^m~12\rlap.{''}8$ and Dec.[J2000]=
$-05^\circ~24'~11''$. 
McMullin et al. (1993) found strong thermal SiO(J=2-1; v=0) emission using 
the Berkeley-Illinois-Maryland Association (BIMA) interferometer. 
This thermal emission was more spatially confined to the 
west of OMC1S showing very broad ($\sim$ 15 km s$^{-1}$) wings, suggesting the 
presence of energetic outflow activity in the region. 

The production of the silicon monoxide molecule (SiO) 
along outflows is mainly attributed to the destruction of the 
dust particles in strong shocks. 
In these violent shocks, the grain-grain collisions
and thermal sputtering inject refractory elements (Si) into the gas 
phase, permitting a quick formation of SiO through reactions involving 
the OH molecule (Schilke et al. 1997; van Dishoeck and Blake 1998).              
Thus, the abundance of the SiO molecules can be enhanced
by several orders of magnitude with respect to the quiescent ambient cloud
values (e. g. Guilloteau et al. 1992).  

Recent surveys of proper motions in the Orion Nebula 
(Bally et al. 2000; O'Dell \& Doi 2003) revealed that at least six relatively 
large Herbig-Haro outflows (HH 202, HH 269, HH 529, HH 203/204, HH 530, 
and possibly HH 528) are originating from a region only a few arcseconds across in 
size located in
OMC1S. O'Dell \& Doi (2003) located this region at 
R.A.[J2000]=5:35:14.56, Dec.[J2000]=-5:23:54.0 
with a positional error of $\pm$1.5$''$, and they referred to
it as the Optical Outflow Source (OOS). 
The relation of these half-dozen Herbig-Haro flows with the molecular outflows found toward
OMC1S has remained uncertain due to the poor angular resolution of the molecular 
observations. 
Moreover, deflections of those optical flows by their parental molecular cloud can be another problem 
in the association of them with molecular flows or exciting objects. Some optical flows 
that have been reported in the literature to be deflected or bent are: 
HH 110 (Reipurth, Raga \& Heathcote 1996),  
and HH34 (Eisl\"{o}ffel \& Mundt 1997; Bally et al. 2006), as have the molecular outflows, 
NGC 2264G (Fich \& Lada 1997), and  NGC1333 IRAS 4A (Choi 2005). 
This mechanism has been successfully modeled by Raga and Cant\'o (1995) and Raga et al. (2002).   
Based on the HH 110 case, they modeled the collision of a supersonic, radiative jet with a dense 
cloud, and found that the spatial velocity difference observed between HH 270 and HH 110 agrees 
surprisingly well with the velocity decrease expected for a radiative jet/cloud collision at 
the deflected angle. Finally, the change in collimation from HH 270 to HH 110 also agrees well
with the broadening of the post-collision jet predicted in the theoretical models.   

In this paper, we present new sensitive and high angular 
resolution SiO J=5$\rightarrow$4; $v=0$ SMA observations 
of the OMC1 South region that reveal a cluster of four hidden compact outflows
 that are present in this region, and which seem to be the base 
of some remarkable Herbig-Haro flows that also emanate from here. 
In $\S$ 2 we describe our SiO observations. 
In $\S$ 3 we report the results of the imaging in SiO thermal emission. 
Finally, a summary is given in $\S$ 4. 

\section{Observations}

The observations were made with the SMA\footnote{The Submillimeter 
Array (SMA) is a joint project between the Smithsonian 
Astrophysical Observatory and the Academia Sinica Institute 
of Astronomy and Astrophysics, and is funded by the Smithsonian
Institution and the Academia Sinica.} during 2004 
October 29 in its "compact" configuration with 6 antennas 
in the array.
The phase reference center of the observations was  
R.A.[J2000]$=05^h~35^m~14\rlap.{''}0$ and Dec.[J2000]$=-05^\circ~24'~00''$.  
The frequency was centered on the SiO(J=5-4 v=0) line at 217.104 GHz in the lower sideband, 
while the upper sideband was centered at 227.104 GHz.
The SMA digital correlator was configured with a band of 128 channels over 104 MHz, 
which provided 0.8125 MHz (1.1 km s$^{-1}$) resolution. 
The zenith opacity ($\tau_{230GHz}$), measured with the National Radio Astronomy Observatory (NRAO) 
tipping radiometer located at the Caltech Submillimeter Observatory (CSO), 
varied between 0.32 and 0.53. 
The system temperature also varied during the run between 350 and 750 K.
The phase and amplitude calibrators were the quasars 0423-013 and 3C120 with measured 
flux densities of 1.80 $\pm$ 0.02 Jy and 0.60 $\pm$ 0.01 Jy, respectively.
The uncertainty in the flux scale is estimated to be 20\%, based on the SMA monitoring of quasars.
Observations of Titan provided the absolute scale for the flux density calibration. 
Further technical descriptions of the SMA are found in Ho, Moran \& Lo (2004).
    
The data were calibrated using the IDL superset MIR originally developed 
for the Owens Valley Radio Observatory (Scoville at al. 1993) and adapted for 
the SMA\footnote{The MIR cookbook by Charlie Qi
 can be found at http://cfa-www.harvard.edu/$\sim$cqi/mircook.html}.
The calibrated data were imaged and analyzed in 
the standard manner using the MIRIAD and AIPS packages. 
We weighted the \sl (u,v) \rm data using the ROBUST parameter of INVERT set to 5, 
optimizing for a maximum sensitivity in each line image. This option is recommended
to achieve the largest signal-to-noise ratio possible, although some angular
resolution is sacrificed. 
The synthesized beam had dimensions of 
$2\rlap.{''}8 \times 1\rlap.{''}7$ (or 1300 AU $\times$ 800 AU at an adopted distance of 460 pc, 
see Bally, O'Dell, \& McCaughrean 2000) with a P.A. = $82.2^\circ$.
The resulting line image rms was 70 mJy beam$^{-1}$.
In this paper
we used LSR radial velocities, and the transformation
adopted to heliocentric radial velocities is
v(heliocentric) = v(LSR) + 15.2 km s$^{-1}$.

\section{The New SiO Outflows} 

Figure 1 shows the contour and color scale images of the 
$^{12}$CO J=2$\rightarrow$1, SiO J=5$\rightarrow$4; $v=0$ 
and 1.3 mm continuum thermal emission detected 
toward the OMC1S region in the Orion Nebula
using the Submillimeter Array. 
The upper panel shows the $^{12}$CO J=2$\rightarrow$1 highly collimated 
($\sim$ $3^{\circ}$) and fast  ($\sim$ $\pm$80 km s$^{-1}$) 
bipolar outflow and the 1.3 mm continuum emission imaged by Zapata et al. (2005). 
The exciting source of this outflow seems to be
the source 136-359, or the infrared source C (Gaume et al. 1998). 
However, the bolometric luminosity of this source appears 
to be far too low to account for the powerful molecular outflow.
We note that this $^{12}$CO bipolar outflow was the only outflow detected 
in those observations. The lower panel shows the four new compact 
 SiO J=5$\rightarrow$4; $v=0$ molecular 
outflows associated with the 
millimeter and infrared sources 136-356, 137-347, and 137-408 
as well as the SiO molecular clumps associated with the redshifted side 
of the collimated $^{12}$CO J=2$\rightarrow$1 flow. 
In Tables 1 and 2, we give their observational and physical 
parameters, respectively. The observed parameters were determined from a linearized 
least-squares fit to a Gaussian ellipsoid function using the 
task IMFIT of AIPS.
    
In what follows, we will discuss each of the four thermal SiO molecular 
outflows and the thermal SiO emission associated with 
the redshifted side of the CO flow, separately.

\subsection{136-356 Bipolar Outflow: The Most Powerful SiO Outflow in the Region}

In Figure 1, in the lower panel, we can see how this strong 
SiO outflow is emanating bipolarly with a west(blueshifted)-east(redshifted) 
orientation from the infrared source 136-356 or source "B" (Gaume et al. 1998;
marked with a green star), 
a luminous ($\sim$ 3 $\times$ 10$^{3}$ L$_{\odot}$) infrared protostar
of spectral type B2 (Gaume et al. 1998). 
We therefore propose that this object is likely the exciting source of
the outflow. 
It is interesting to note that this source appears to be a relatively 
evolved protostar, possibly a class I or maybe class II object (Andr\'e et al. 1993). 
This classification is due to the fact that it shows very weak 1.3 mm continuum emission. 

This outflow shows an extremely broad range of velocities: 
in its blueshifted component it displays
gas velocities in the range of about -80 to -20 km s$^{-1}$, 
while in its redshifted component the range is +105 to +25 km s$^{-1}$ 
with respect to the ambient cloud velocity,  
which is taken to be 5 km s$^{-1}$ (see Figure 3 and 4). 
The position angle of the whole outflow (redshifted and
blueshifted lobes) can be determined
with great accuracy: 108.4$^{\circ}$ $\pm$ 0.1$^{\circ}$, and 
its collimation angle is $\sim$ 13$^{\circ}$. This small collimation
angle is consistent with an extremely young source (Arce \& Sargent 2006).
With our present angular resolution 
($2\rlap.{''}8 \times 1\rlap.{''}7$), 
we can resolve the redshifted lobe, while the blueshifted lobe 
was only resolved in the east-west orientation (the major axis). We show their deconvolved 
parameters in Table 1. 

We tried to test if the millimeter continuum source at the
center of this outflow is elongated perpendicular to the outflow axis (as expected
in the case of a circumstellar disk, but the continuum source is relatively weak and a
reliable deconvolution of its dimensions was not possible. 

The spatial structure of the high and low velocity gas of the bipolar outflow as a 
function of the radial velocities (PV diagram) along the major axis is shown in Figure 4. 
This image shows that there is a tendency of the SiO thermal emission to be located very close to 
the proposed exciting object. 
Moreover, in its redshifted component, the line width is extremely broad, over 70 km s$^{-1}$,
as measured at the lowest displayed intensity contours.
Similarly, the linewidth on the blueshifted side is $\sim$ 50 km s$^{-1}$.
We suggest that a wide-angle wind may be the 
driving mechanism                                        
of the 136-356 bipolar outflow (see, Shu et al. 1991 and Shang et al. 1998).  
Palau et al. (2006) found a PV structure in their observations of the HH 211 bipolar
outflow which is very similar 
to our Figure 4 (see the spatial distribution of the SiO in their Figure 3). 
The observed wide range of velocities at the base of outflow is attributed to a protostellar wind
with a large opening angle, thereby yielding a maximum spread of velocities.
A pure, highly collimated jet model, in which the
velocity vectors point only 
in the polar direction, cannot produce the broad line width,  as explained in 
Palau et al. (2006). Furthermore, Arce and Goodman (2002) also predict this kind 
of structure on their PV diagram for a molecular outflow driven by a wide-angle wind.
However, a very turbulent outflow, even if highly collimated could
also explain the observations. If we assume that the broad emission from the outflow lobes
is produced by a wide-angle wind, we can obtain a crude estimate of the total
space velocity of the outflow as follows. Under this assumption, the velocity width
of the outflow is given by $\Delta v \simeq 2~v_{TOT}~sin(\theta_{coll}/2)$, where
$v_{TOT}$ is the space velocity of the outflow and $\theta_{coll}$ is the  
collimation angle of the outflow. Adopting $\Delta v \simeq 50~km~s^{-1}$
and $\theta_{coll} \simeq 13^\circ$, we obtain $v_{TOT} \simeq 220~km~s^{-1}$,
which is a plausible value.
 
From Figure 6, we note that the easternmost clump seems 
not to form part of this strong bipolar outflow. 
This component is located $\sim$ 10$''$ away from the exciting object and shows low velocity 
gas (10 to 30 km s$^{-1}$), unlike the extremely high velocity gas observed in 
this outflow. If this easternmost clump was ejected by 
136-356, the dynamical age would be of order
1,300 years, much older than the age of $\sim$200 years of the
compact SiO outflow. A possibility is that this thermal emission is tracing another outflow,
perhaps powered by the source 141-357 only a few arcseconds away (see Figure 1). 
The nature of this molecular component is unclear.         

As mentioned in the introduction, the HH269 flow seems to be originating from a 
region only a few arcseconds across located in OMC1S. The HH 269 object 
is a blueshifted flow located due west of the OMC1S core and 
moves toward P.A. = +255$^{\circ}$, with a proper motion of about 50 km s$^{-1}$ 
(Bally, O'Dell \& McCaughrean 2000).
O'Dell \& Doi (2003) deduced that the expected location of the 
driving source responsible for the HH 269 flow should be located in the OOS,
where no radio (Zapata et al 2004a,b), 
infrared (Robberto et al. 2005) or X-ray (Grosso et al. 2005) sources are detected. 
We have marked with a cross the position of the OOS region in the Figures 1 and 2.
In Figure 2, we have overlapped the SiO velocity-integrated intensity detected by us with the
H$\alpha$ emission image of O'Dell \& Wong (1996) from the Hubble Space Telescope.
In this image we can see how the blueshifted side of the 136-356
bipolar SiO outflow is roughly aligned with the blueshifted HH 269 object,
the differences of alignment could be explained by the jet deflection
mechanism modeled by Raga and Cant\'o (1995) and discussed in the
introduction.
Moreover, both optical and radio flows display
high velocity features. Then, it is tempting to propose that the source B or 136-356 might be 
energizing this optical large-scale object. 
However, there are strong arguments against this identification and we refer the
reader to Henney et al. (2006) for a detailed discussion. 

In this paper, we have estimated the kinematic ages of the SiO outflows
by dividing the projected displacement in the
plane of the sky over the radial velocity. In the case of this outflow,
the age comes to be of order 200 years (see Table 2). We note that this manner of estimating 
kinematic ages can be very uncertain. An object moving at velocity $v$ forming an angle $\theta$
with respect to the line of sight, will have a line-of-sight radial velocity
of $v_r = v~cos(\theta)$. On the other hand, the projected displacement 
in the plane of the sky will be $d_p = d~sin(\theta)$, where $d$ is the real displacement.
Then, $t'$, the age estimated by this method, is $t' = d_p/v_r = 
t~ ctn(\theta)$, where $t=d/v$ is the real age. If 
$\theta \simeq 45^\circ$, then $ctn(\theta) \simeq 1$
and the estimated age is close to the real value. However, if the velocity vector
is either close to the line-of-sight or to the plane of the sky, the error can
be very large, reaching factors of about 5 for the cases of $ctn(\theta) = 10^\circ$ or
$ctn(\theta) = 80^\circ$. Then, the estimated ages should be taken with caution.

Only a few arcseconds east of this strong SiO outflow 
there is another remarkable Herbig-Haro that appears to be emanating from the 
OOS region, HH 529 (see Figure 2). The HH529 object consists of a chain 
of blueshifted bow shocks and knots moving toward the east at P.A. = -260$^{\circ}$ 
at a declination of about -5$^{\circ}$24$'$ and reaching 
spatial motions of up to 110 km s$^{-1}$ (Bally, O'Dell \& McCaughrean 2000).
A few sources have been proposed to power the HH 529 object:
the infrared source TCC 009 (McCaughrean \& Stauffer 1994; 
proposed by Bally, O'Dell \& McCaughrean 2000),
the infrared sources A, B, and C (Gaume et al 1998; 
proposed also by Bally, O'Dell \& McCaughrean 2000), 
the OOS region O'Dell and Doi (2003), and the source IRS2 of Smith et al. (2004). 
Possibly, the best candidate for the exciting source 
of this bow shock is the object TPSC-1 (Lada et al. 2000), which is seen only in the L band.
We have indicated the position and motion of this HH object in Figure 2.
>From this image it is evident that the HH 529 object does not align
accurately with the 
redshifted lobe of this new SiO outflow (136-356). However, following again 
the jet deflection mechanism explained in the first section, we speculate that 
this molecular component, originally redshifted, 
may have been deflected due to an interaction with the molecular cloud and became 
the blueshifted optical object HH 529. These
possible associations are discussed in a following paper (Henney et al.
2006).  
 
\subsection{Other SiO Molecular Flows}

\subsubsection{137-347 Monopolar Outflow}

This northwest-southeast monopolar SiO outflow seems to emanate from 
the millimeter source 137-347 (see Figure 1), 
a source with no counterpart at infrared wavelengths (see Lada et al. 
2000, 2004; Muench et al. 2002, Smith et al. 2004, Robberto et al. 2005), 
perhaps suggesting that it could be associated with a class 0 protostar.   
We propose that the object 137-347 is possibly its exciting source. 
The 137-347 monopolar outflow shows redshifted emission of moderate  
gas velocity ($\sim$ 5 to 31 km s$^{-1}$) with respect to the ambient 
cloud velocity  (see Figures 3 and 5). 
The position angle of this outflow is 161.1$^{\circ}$ $\pm$ 0.5$^{\circ}$. 
The spatial distribution of the integrated SiO J=5$\rightarrow$4 thermal emission shows 
that the source is not resolved  along its minor axis, and this is consistent with a 
molecular clump of gas moving at moderate velocity. 
At present it is not possible to establish if 
its monopolarity is due to anisotropic ambient cloud conditions or to
an intrinsic asymmetry in the flow.

The PV diagram of this flow along the major axis is shown in Figure 7. 
This image shows
the strongest SiO molecular emission at about
23 km s$^{-1}$ and a distance from the exciting source of $\sim$900 AU.
Analyzing this image, we note that this outflow also appears to be driven 
by a wide-angle wind starting
close to its exciting object, since it shows a broad range of velocities
very close to the exciting object, as observed in the bipolar outflow 136-356.

The 10 to 30 km s$^{-1}$ clump discussed in section 3.1, that we tentatively  
proposed as being excited by the continuum source 141-357, falls close to the
direction defined by the 137-347 monopolar outflow and could potentially be
excited by the latter source. Proper motion studies of these outflows are
needed to identify more firmly the exciting sources.

\subsubsection{137-408 Multipolar Outflow}

\subsubsubsection{137-408 Bipolar Outflow: East-west Flow}

This SiO bipolar outflow is found to be very young 
(with a dynamical age of about 200 yrs), exhibiting high velocity and 
well centered on the strong millimeter source 137-408 (see Figure 1).   
We note that the source 137-408 does not have centimeter or infrared counterparts
(see Lada et al. 2000, 2004; Smith et al. 2004; Robberto et al. 2005, Zapata et al. 2004a,b),
possibly because it is a genuine embedded class 0 object.  We propose that 137-408 is 
likely the exciting source of the outflow.

The 137-408 bipolar outflow shows redshifted and blueshifted gas velocity
at about $\pm$70 km s$^{-1}$ quite symmetric 
with respect to the ambient cloud velocity (see Figures 3 and 8). 
The position angle of this object is 129.3$^{\circ}$ 
$\pm$ 0.9$^{\circ}$, and its collimation angle is $\sim$ 15$^{\circ}$. 
The northwest lobe is resolved only along its major axis, and we show its 
deconvolved parameters in Table 1. 
Its PV diagram along the major axis is shown in Figure 8. Analyzing this image we find that 
most of the blueshifted SiO thermal emission falls in a range of velocities between 
-30 to -80 km s$^{-1}$, while the redshifted
emission falls between 30 to 80 km s$^{-1}$.

From our previous millimeter continuum image of the OMC1S region (Figure 1), we believe that 
FIR 4 and CS-3 sources are likely the same source (this
possibility was first proposed by Gaume et al. 1998), and that this source 
could be related with our strong millimeter source 137-408, only a few arcseconds 
from both sources. However, given the uncertainties in the available
positions for FIR 4 and CS-3, we cannot rule out the
possibility that CS-3 is associated with 137-408 and that
FIR 4 is associated with the continuum structure
located about 2$''$ to its west. 
This outflow can be related to one of the HH objects of east-west
orientation that emanate from FIR 4 or CS-3; however, there is no clear
relation to them.

\subsubsubsection{137-408 Monopolar Outflow: South Flow}

This SiO outflow also appears to emanate from the millimeter source 137-408 (see Figure 1). 
Therefore, in the same way as the previous 
SiO flow, the object 137-408 could be 
its exciting source. The 137-408 south flow shows redshifted emission 
of low (0 to 27 km s$^{-1}$, for the most southern component) and moderate 
(29 to 56 km s$^{-1}$, for the component nearest to 137-408) gas velocity with respect 
to the ambient cloud velocity (see Figure 9). The position angle of this 
outflow is $\sim$ 34$^{\circ}$. 
The spatial distribution of the integrated SiO thermal emission of the nearest 
component to the source 137-408 is found to be angularly unresolved, 
while the southern component is angularly resolved and 
we give its deconvolved parameters in Table 1.
 
We note that this flow shows a different morphology as compared with the other molecular 
outflows. It shows a clumpy structure along its major axis with every clump or "bullet" 
displaying a different range of velocities (see Figure 9). Moreover, it is the 
most extended outflow detected by us, being $\sim$ 15$''$ long
and its dynamical age is an order of magnitude larger than the other
flows discussed here (see Table 2).

When comparing our observations with those of the $^{12}$CO J=2$\rightarrow$1 
observations of Schmid-Burgk et al. (1990) 
toward the OMC1S region, we find that
they detected a 120$''$ long redshifted monopolar low velocity 
(15 km s$^{-1}$) outflow with a position angle of 31$^{\circ}$, presumably ejected 
from the sources FIR 4 or CS-3. As discussed above, the (sub)millimeter sources 
FIR 4 and CS-3 could be associated with our source 137-408, implying
that this SiO monopolar outflow is emanating from the same source as the CO outflow
and at a P.A. of about 30$^{\circ}$. Moreover these SiO and CO
outflows show low velocity features.
Hence, we propose that the SiO outflow detected here is the base of
the large-scale CO flow detected by Schmid-Burgk et al. (1990). However, 
we believe that further, more 
sensitive SiO observations of low angular resolution are 
required to confirm this relation.
          
We note that the presence of two outflows associated with 137-408 may indicate 
more than one underlying exciting source. These could be unresolved or one may be 
very faint.

\subsection{SiO Thermal Emission Associated with the Powerful
Southeast-Northwest CO Outflow}

 In Figure 1, we can see that there are
 two molecular clumps (labeled as S1 and S2) associated with the 
 redshifted lobe of the highly collimated northwest-southeast CO bipolar outflow 
 (see Figure 1, upper panel). The S1 clump shows moderate redshifted gas 
(10 to 25 km s$^{-1}$), while the S2 clump shows gas motions at high velocities 
 (40 to 100 km s$^{-1}$), see Figures 3 and 8. 
 We note that the clump S2 shows a little positional offset toward the
 northeast direction of the 
 possible exciting object (source 'C' or 136-359). 

 The southeastern clump (S1) is also very compact, and it is located 
 in one of the knots of the southeast-northwest redshifted CO lobe (see Figure 1). 
 As in the other outflows discussed here,
 we suggest that this SiO clump is produced by 
 an internal shock in the flow,  where the abundance of the SiO molecules can be enhanced 
 by several orders of 
 magnitude with respect to the quiescent ambient cloud by the destruction of  
 dust particles (L1157; Gueth, Guilloteau and Bachiller 1998;
 L1448; Dutrey et al. 1997).   

 The spatial structure of the high and low velocity gas of 
 the S1 and S2 clumps as a function of the radial velocities 
 (PV Diagram) along a P.A. of 110$^{\circ}$ is shown 
 in Figure 10. 
 
\subsection{Physical Parameters of the SiO Outflows}
 
 Following Liu et al. (1998), we calculate the physical parameters of all
 the new  molecular outflows, e.g. masses, momenta, energies 
 and mechanical luminosities, and the results are presented in Table 2.
 We assumed LTE conditions and optically thin emission, a rotational temperature of 80 K, 
 and an abundance ratio of SiO/H$_2$ equal to 1 $\times$ 10$^{-7}$ (Ziurys \& Friberg 1987; Mikami 
 et al. 1992; Zhang et al. 1995). 
It is important to mention that this ratio can change in every molecular outflow
 and the masses can be over/underestimated.  
 We also took the major and minor axes of the synthesized Gaussian beam  
 to be 2.8$''$ and 1.7$''$, respectively, 
 the upper energy level for the SiO J=5$\rightarrow$4 transition to be E$_{u}$=14.48 K, a 
 rotational constant for SiO of B=21,787 MHz and a dipole moment of 3.1 Debyes
 (Liu et al. 1998).

 From Table 2 we note that the implied kinematic luminosity for 
 the strongest SiO molecular outflow should be about 450 L$_{\odot}$, 
 a factor of $\sim$ 8 less than the total bolometric luminosity ($\sim$ 3 $\times$ 10$^3$ L$_{\odot}$) 
 of the infrared source 'B'. Even when the situation in this case is not
 as critical as in the CO outflow reported by Zapata et al. (2005), where the
 kinematic luminosity of the outflow seems to \sl exceed \rm the bolometric luminosity
 of the exciting source, it is still very dificult to explain the large
 kinematic luminosity of this SiO outflow. Statistically, the
 kinematic luminosity of outflows is of order of only
 0.001-0.01 of the bolometric luminosity of the exciting source (Wu et al. 2004).
 One possibility is that
 the SiO is greatly enriched in this flow, leading to an overestimate
 in the total mass of the outflowing molecular gas. 
 
\section{General Discussion}
       
 We note that most of the SiO outflows are compact ($<$ 3500 AU  $\times$ $<$ 1500 AU),
 located close to their proposed exciting objects, and display a high dispersion of velocities. 

 It is puzzling that in our previous CO observations we
 did not detect these three monopolar and bipolar SiO flows. 
 A possible explanation for the non-detection of these new SiO outflows in the CO data would
 be that two of them are low velocity outflows. The 137-408 monopolar south outflow (0 to 27
 km s$^{-1}$ for the southern component, and 30 to 50 km s$^{-1}$ for the northern component) and
 the 137-347 monopolar south-east outflow (5 to 35 km  s$^{-1}$), may be confused with extended 
 CO J=2$\rightarrow$1 emission at 
 ambient velocities (-25 to 21 km s$^{-1}$), see Zapata et al. (2005). 
 However, for the other two east-west 
 SiO outflows (136-356 and 137-408) it is not clear why they were not detected as 
 they have extremely high velocities ($>$ 70 km s$^{-1}$), and are apparently massive outflows.
In particular, for the strong SiO outflow associated with 136-356 we estimate, from
comparison with the data of Zapata et al. (2005), that an enhancement
of SiO of order 50 with respect to the adopted value of [SiO/H$_2$]$\simeq$1 $\times$ 10$^{-7}$
is needed in order to explain the non detection of this SiO outflow in CO. 
 
 We also note that the blueshifted velocity component of the highly collimated CO outflow
 shows a faint extension between the second and third molecular "bullets", to the west of
 these components (see Figure 1). If one searches at the position of this weak CO feature in the
 SiO velocity-integrated intensity image, we find
 that this component coincides very well in position with the blueshifted
 side of the 136-356 bipolar outflow. Therefore, in the CO velocity-integrated intensity image, 
 we may be seeing weak CO emission associated with the blueshifted component of 
 the 136-356 SiO bipolar outflow.
 In such a way, this extension can be the key to explain the 
 nature of those two new outflows. Probably, 
 the SiO emission is tracing high density entrained gas with a broad range of velocities and 
 closer to their exciting objects, associated perhaps with their primary jets 
 (see Hirano et al. 2006).
 The CO emission, is possibly tracing low density entrained gas, with a less broad 
 range of velocities as has been observed by Palau et al. (2006) in the HH211 flow. 
 Following this premise, in these SiO observations, we have imaged unresolved 
 emission from the jet itself, as observed by Hirano et al. (2005) and 
 Palau et al (2006). However, for this mechanism to work, we need strong chemical mechanisms 
 that would strongly favor the detection of each molecule in the two components of
 different density. 
 
 At this point, it is not well understood yet why the SiO is a selective tracer. 
 In some cases it seems to be associated with the most excited parts of the flows 
 (136-356 and 137-408 outflows; this paper, and HH211; Hirano et al. 2006; Palau et al. 2006) 
 while in other cases it is associated with internal shocks in the outflows,  
 (L1157; Gueth, Guilloteau and Bachiller 1998, L1448; Dutrey et al. 1997, 
 and the 136-359 CO outflow; this paper). One possibility for unifying these two mechanisms would 
 be if the SiO molecule is formed/excited in shocks that are propagating down the jet 
 ("internal working surfaces"), as proposed by Arce et al. (2006). 
 
In Table 3 we show a list of tentative associations between the HH objects, the molecular outflows, 
and their proposed exciting sources located in the OMC1S region.

\section{Summary}

 We have observed the OMC1 South region in the SiO J=5$\rightarrow$4 $v=0$
 molecular line with high 
 angular resolution ($2\rlap.{''}8 \times 1\rlap.{''}7$). We found for the first time  
 a cluster of four hidden compact bipolar and monopolar outflows that show high, 
 moderate and low velocity gas and appear to be energized by millimeter 
 and infrared sources associated with the OMC1 South region. Furthermore, these SiO 
 outflows seem to be the base of some remarkable Herbig-Haro flows that  
 emanate from this region. The specific results and conclusions of 
 this study are as follows:

1. We found four compact molecular outflows:
 
\begin{itemize}
\item  A strong east-west bipolar molecular outflow 
       that appears to be powered by the luminous ($\sim$ 3 $\times$ 10$^3$ L$_{\odot}$) infrared source 
       'B' or 136-356. This molecular flow is very compact ($<$ 3600 AU) and shows a quite broad 
       range of velocities ($\sim$ $\pm$ 100 km s$^{-1}$). By analyzing 
       its PV diagram, we find evidence that its redshifted component is been driven 
       by a wide-angle wind. This blueshifted component 
       could be powering the HH 269 object, if we assume that this molecular
       outflow is being deflected. This and other possible associations in the region
       are discussed in detail in Henney et al. (2006). 

 \item A north-south compact monopolar redshifted outflow associated with the source 
       137-347 and has gas with velocities ranging between $\sim$ 5 and 31 km s$^{-1}$.
       The spatial distribution of the integrated SiO J=5$\rightarrow$4 thermal emission is 
       unresolved and is consistent with a molecular clump of moderate velocity. By 
       analyzing its PV diagram, we also find evidence that it is being driven by 
       a wide-angle wind.
 
 \item A multipolar outflow associated with the millimeter source 137-408. One of the components 
       is emanating almost in an east-west orientation, while the other one is emanating 
       in the southeast-northwest direction. The east-west bipolar outflow is very young (about 200 yrs), 
       shows a broad range of velocities ($\sim$ $\pm$ 80 km s$^{-1}$), and 
       has a P.A. of 129.3$^{\circ}$.  
       On the other hand, the northwest-southeast monopolar flow shows gas at moderate velocities ranging 
       between $\sim$ 3 and 60 km s$^{-1}$ and it has a
       position angle of 34$^{\circ}$. Comparing 
       with the orientation (31$^{\circ}$) and the integrated gas velocity ($\sim$ 15 km $s^{-1}$) 
       of the red-shifted component of the large scale CO molecular outflow that emanates from this region 
       (Schmid-Burgk et al. 1990), we propose that the monopolar SiO flow is the 
       base of this CO flow. 
       However, we believe that further sensitive SiO observations of low angular resolution are
       required to confirm this relation. Finally, we also note that the presence of two 
       outflows associated with 137-408 may indicate more than one underlying exciting source.

\end{itemize}

2. We found two compact SiO molecular clumps S1 and S2 associated with the
redshifted component of the CO outflow (Zapata et al. 2005). The northwestern 
component (S2) is very compact
and its position is almost coincident
with the source 136-359, the proposed exciting source of the CO flow.
The southeastern clump (S1) is also very compact,
shows moderate redshifted gas velocity (10 to 25 km s$^{-1}$) and it is located in one of
the knots of the 
southeast-northwest CO outflow. We therefore suggest that this molecular SiO clump is produced by 
an internal shock in the flow,  where the abundance of the SiO molecules can 
be enhanced by several orders of 
magnitude with respect to the quiescent ambient cloud value by destruction of the 
dust particles.   
 
3. We find that the SiO molecular outflows are compact ($<$ 3500 AU  $\times$ $<$ 1500 AU), 
show a broad range of velocities (-80 to +110 km s$^{-1}$), 
and in most of the cases, are located very close to their exciting sources. We propose  
that the SiO thermal emission is tracing the youngest and most highly excited parts of some 
outflows which cannot be detected in other molecules. Moreover, the SiO line
 observations can detect flows that in other
molecular tracers appear confused with the ambient velocity cloud emission.   
 
\acknowledgments  
We thank all the SMA staff members for making these observations possible.
LAZ acknowledges the Smithsonian Astrophysical Observatory 
for a predoctoral fellowship. 
LFR thanks CONACyT, M\'exico and DGAPA, UNAM for their support.

\bibliographystyle{apj}

\clearpage

\begin{deluxetable}{l c c c c c }
\tablecolumns{6}
\tabletypesize{\scriptsize}
\tablewidth{0pc}
\tablecaption{Observational Parameters of the SiO Thermal Emission}
\tablehead{
\colhead{}                       &
\colhead{}                       &
\colhead{}                       &
\multicolumn{3}{c}{Parameters} \\
\cline{4-6}
\colhead{Outflow }                &
\multicolumn{2}{c}{Coordinates}  &
\colhead{Deconvolved}           &
\colhead{} \\
\colhead{Name} &
\colhead{$\alpha_{2000}$}  &
\colhead{$\delta_{2000}$}  &
\colhead{Size} &
\colhead{P.A.}  &
}
\startdata
137-347             & 05 35 13.740 & -05 23 48.79 &  2.1 $\pm$ 0.4 $\times$ -  &  161.1$^{\circ}$ $\pm$ 0.5$^{\circ}$ \\
136-356 Redshifted  & 05 35 13.817 & -05 23 56.79 & 3.006$''$ $\pm$ 0.002$''$ $\times$ 0.35$''$ $\pm$ 0.01$''$  &  106.96$^{\circ}$ 
$\pm$ 0.08$^{\circ}$  \\
137-356 Blueshifted & 05 35 13.250 & -05 23 54.30 & 5.84$''$ $\pm$ 0.2$''$ $\times$ --  &  109.78$^{\circ}$ $\pm$ 0.07$^{\circ}$  \\
S1                  & 05 35 14.035 & -05 24 02.47 &   -  &  -   \\
S2                  & 05 35 13.511 & -05 23 59.42 &   -  &  -  \\
137-408 Redshifted  & 05 35 13.946 & -05 24 08.16 & 1.00$''$ $\pm$ 0.02$''$ $\times$ --   & 129.3$^{\circ}$ $\pm$ 0.9$^{\circ}$    \\
137-408 Blueshifted & 05 35 13.534 & -05 24 06.43 &   -  &  -  \\
137-408 (South-1)   & 05 35 13.662 & -05 24 10.51 &   -  &  -  \\
137-408 (South-2)   & 05 35 13.416 & -05 24 15.48 & 1.89$''$ $\pm$ 0.03$''$ $\times$ --  &  49.3 $^{\circ}$ $\pm$ 0.8$^{\circ}$  \\
\enddata
\tablecomments{\footnotesize Units of right ascension are hours, minutes, and seconds, and
units of declination are degrees, arcminutes. }
\end{deluxetable}

\clearpage

\begin{deluxetable}{l c c c c c c c}
\tablecolumns{8}
\tabletypesize{\scriptsize}
\tablewidth{0pc}
\tablecaption{Physical Parameters of the SiO Molecular Outflows}
\tablehead{
\colhead{Outflow} &
\colhead{Projected Size} &
\colhead{Dynamical Age}     &
\colhead{Mass}                &
\colhead{Momentum}            &
\colhead{Energy}                       &
\colhead{Mechanical Luminosity}        \\
\colhead{Name} &
\colhead{[AU]} &
\colhead{[yr$^{-1}$]} &
\colhead{[M$_{\odot}$]}  &
\colhead{[M$_{\odot}$ km s$^{-1}$]}  &
\colhead{[10$^{46}$ erg]} &
\colhead{[L$_{\odot}$]}  &
}
\startdata
137-347              & 970  & 188  & 0.05    & 1       &  0.025  & 2.4\\
136-356 Redshifted   & 1400 & 160  & 0.80    & 45      &  2.5    & 400\\
136-356 Blueshifted  & 2700 & 192  & 0.19    & 7.5     &  0.3    & 50 \\
S1                   & -  &  -   & 0.01    & 0.15    &  0.025  &  - \\
S2                   & - &  -   & 0.04    & 2       &  0.1    &  - \\
137-408 Redshifted   & 460 & 107  & 0.02    & 1       &  0.025  & 8.8\\
137-408 Blueshifted  & - &  -   & 0.005   & 0.2     &  0.005  & 1.8\\
137-408              & 900 & 3000 & 0.04    & 1       &  0.025  & 0.3\\
\enddata
\tablecomments{\footnotesize To calculate energy and momentum, we use as the typical
velocity one half the highest velocity displayed by the molecular outflows.}

\tablecomments{\footnotesize To calculate those physical parameters we assumed a fractional abundance between 
SiO and H$_2$ equal to 1 $\times$ 10$^{-7}$, found by Ziurys \& Friberg (1987), Mikami et al. (1992), and Zhang et al. (1995) toward the Orion-KL and L1157 molecular ouflows, respectively.}
\end{deluxetable}

\clearpage

\begin{deluxetable}{l c c c}
\tablecolumns{4}
\tabletypesize{\scriptsize}
\tablewidth{0pc}
\tablecaption{Tentative relation between the Herbig-Haro Objects and Molecular Outflows that emanate from OMC1 South.}
\tablehead{
\colhead{Herbig-Haro}                   &
\colhead{Molecular}                     &
\colhead{Proposed Exciting}                      &
\colhead{References}                   \\
\colhead{Object}                        &
\colhead{Outflows}                      &
\colhead{Sources}                       &
\colhead{}                             
}
\startdata
HH 530  & 137-408 south outflow?        &  137-408  or FIR4/CS3?    &  BAL00; this paper \\
HH 625  & 136-359 bipolar outflow (NW)  &  136-359  or source C    & OD03; ZA05 \\
HH 528  & 136-359 bipolar outflow (SE)? &  136-359  or source C?    & This paper; OD03 \\
HH 269  & 136-356 bipolar outflow (W)?  &  136-356  or source B?    & This paper  \\   
HH 529  &  No association               &  OOS                     & OD03  \\
HH 202  &  No association               &  143-353; Smith-IRS1     & ZA04; SM04\\
HH 203+HH 204 & No association          &  143-353; Smith-IRS1     & ZA04; SM04\\
\enddata
\tablecomments{\footnotesize OD03=O'Dell and Doi (2003); SM04=Smith et al. (2004); ZA04= Zapata et al. (2004b); 
ZA05=Zapata et al. (2005); BA00=Bally, O'Dell \& McCaughrean (2000)}
\end{deluxetable}

\clearpage

\begin{figure}
\begin{center}\vspace{-0.5cm}\hspace{0.00cm}
\includegraphics[ angle=-90, width=.5\textwidth]{f1a.eps}
\includegraphics[ angle=0, width=.5\textwidth]{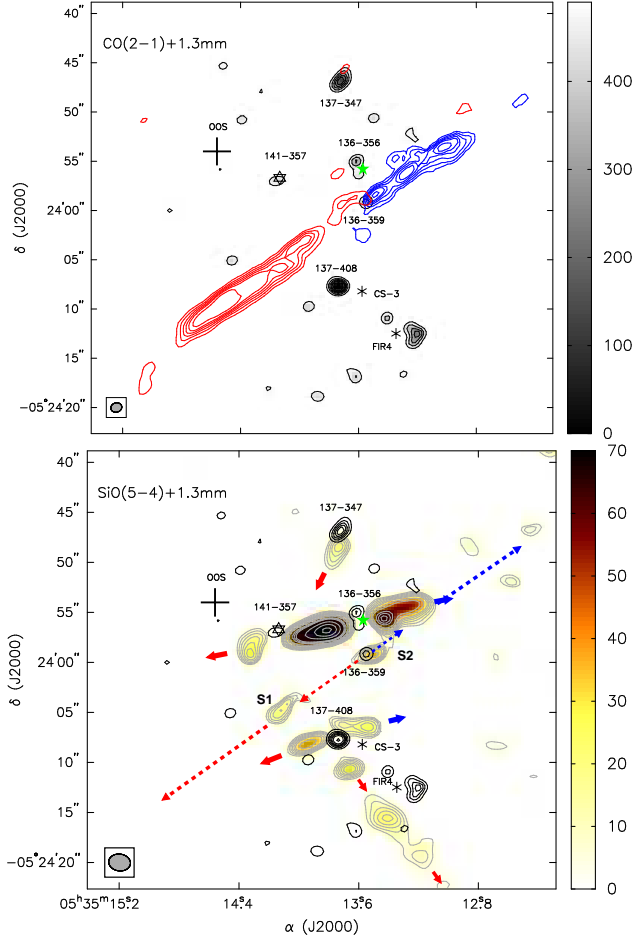}\vspace{-0.2cm}
\caption{{\it Top)} SMA CO J=2$\rightarrow$1 velocity-integrated intensity
(blue and red contours) and
1.3 mm  continuum image (black contours and grey scale) towards the OMC1S region. 
The CO contours are 3, 6, 9, 12, 15, 18, 20 and 30 times 1.45 Jy beam$^{-1}$ km s$^{-1}$, 
the rms noise of the velocity-integrated intensity 
image. The velocity range for the blueshifted gas is
from $-$80 to $-$26 km s$^{-1}$ and that for the redshifted gas is
from +22 to +82 km s$^{-1}$. The scale bar indicates the 1.3 mm continuum emission 
in mJy beam$^{-1}$.
This image was taken from Zapata et al. (2005).  
{\it Bottom)} SMA SiO J=5$\rightarrow$4 velocity-integrated intensity 
(grey contours and yellow scale)
and 1.3 mm continuum (black contours)  emission toward the same region.
The 1.3 mm  continuum image is the same shown in the top panel.
The SiO contours are 3, 4, 5, 6, 7, 8, 10, 12, 16, 18, 20, 30, 40 and 50 $\times$ 1.4
 Jy beam$^{-1}$ km s$^{-1}$, the rms noise of the velocity-integrated intensity image.
The integration is over the velocity range $-$77 to 103 km s$^{-1}$.
The continuous arrows indicate the blueshifted/redshifted SiO thermal emission.
The dashed arrow displays the position of the CO collimated outflow.  
The scale bar gives the SiO line integrated emission in Jy beam$^{-1}$ km s$^{-1}$.
The continuum contours are  3, 6, 9, 12, 15, 18, 20, and 30 $\times$ 11 mJy beam$^{-1}$,
the rms noise of the continuum image in both panels. 
The half power contour of the synthesized beam for
each molecular transition is shown in the bottom left corner of each image.
In both images, the green star indicates the position of the source B or 
source 136-356  (Gaume et al. 1998; Zapata et al. 2004b). 
The asterisks, shown also in both panels, denote the position of the millimeter 
sources FIR 4 (Mezger et al. 1990) and CS-3 (Mundy et al. 1986).
The six-pointed star denotes the position of the source 141-357 (Zapata et al. 2005b) }
\label{fig1}
\end{center}
\end{figure}

\clearpage

\begin{figure}
\begin{center}
\includegraphics[scale=1.3, angle=90]{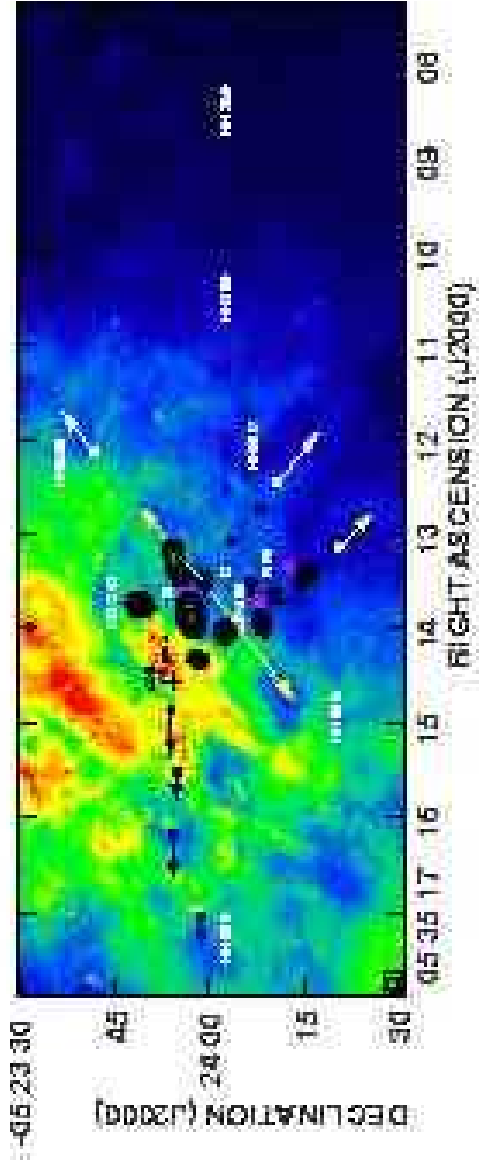}
\caption{SMA SiO J=5$\rightarrow$4 velocity-integrated intensity 
image of the OMC1S region, 
superposed on the H$\alpha$ image of O'Dell \& Wong (1996). The integration is over 
the velocity range of  -77 to 103 km s$^{-1}$. The continuum contours are  
2, 4, 6, 8, 10, 12, 14, 16, 18, 20, 30, 60, 120 $\times$ 11 mJy beam$^{-1}$, 
the rms noise of the continuum image. The half power contour of the 
synthesized beam is shown in the bottom left corner of the image.  
The purple stars indicate the position of the presumed exciting sources
of the SiO molecular outflows (137-347, B, C, and 137-408, 
see Gaume et al. 1998; Zapata et al. 2004b; Zapata et al. 2005). 
The purple asterisk marks the position of the infrared source TPSC-1
(Lada et al. 2000).
The position of the OOS (Optical Outflow Source, O'Dell \& Doi 2003) is 
indicated with a cross. 
The triangle indicates the position of the millimeter source FIR4 (Mezger et al. 1990). 
The yellow arrows indicate the CO molecular outflows that are emanating 
from this zone found by Rodriguez-Franco et al. (1999a) and Schmid-Burgk et al. (1990). 
The white arrows indicate the orientation of the Herbig-Haro flows 530 and 625 
(O'Dell \& Wong 1996;
Bally, O'Dell, \& McCaughrean 2000). The blue arrows indicate the strong O[III] emission detected 
in the Fabry-P\'erot images of  Smith et al. (2004) of the object HH529.
The positions of the HH objects 528 and 269 (O'Dell \& Doi 2003) are also displayed.
 }
\label{fig2}
\end{center}
\end{figure}
\clearpage

\begin{figure}
\begin{center}
\includegraphics[scale=1]{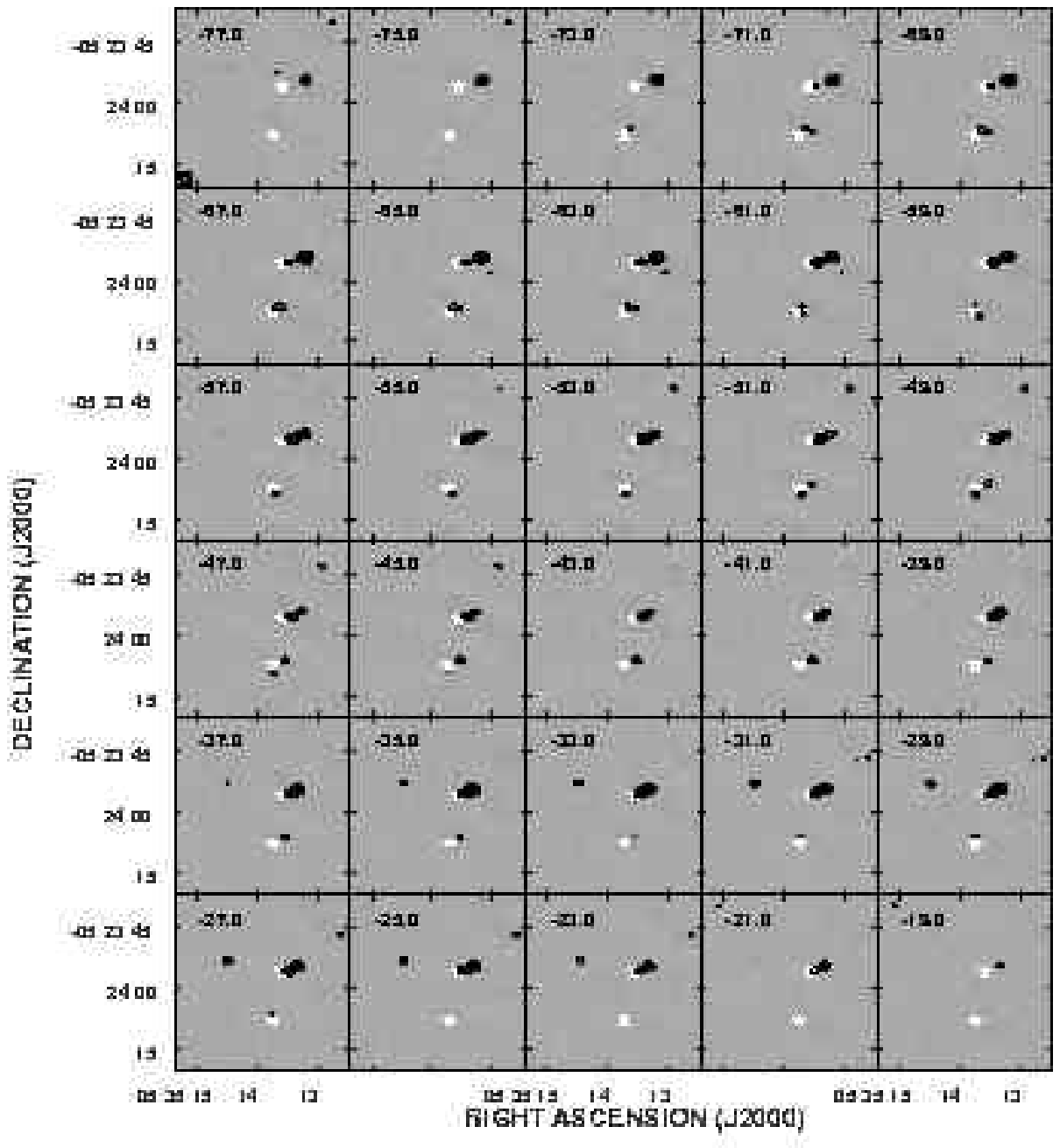} 
\caption{                    
Channel image of the SiO J=5$\rightarrow$4  blueshifted emission of the molecular outflows. 
The emission is averaged in velocity bins of 2 km s$^{-1}$. 
The central velocity is indicated in the upper right-hand corner of each panel 
(the systemic velocity of the ambient molecular cloud is about 5 km s$^{-1}$). 
The FWHM of the synthesized beam is shown in the lower left-hand corner of the first panel. 
The contours are 2, 3, 4, 5, 6, 7, 8, 9, 10, 20, 30 $\times$ 0.07 Jy beam$^{-1}$
km s$^{-1}$, 
the rms noise of the image. The white stars indicate the position of the presumed exciting sources
of the SiO outflows.
The upper white star denotes the position of the source B or source 136-356  
(Gaume et al. 1998; Zapata et al. 2004b), while the lower star denotes the position 
of the source 137-408 (Zapata et al. 2005).
 }
\label{fig3}
\end{center}
\end{figure}

\clearpage


\begin{figure}
\begin{center}
\includegraphics[scale=1]{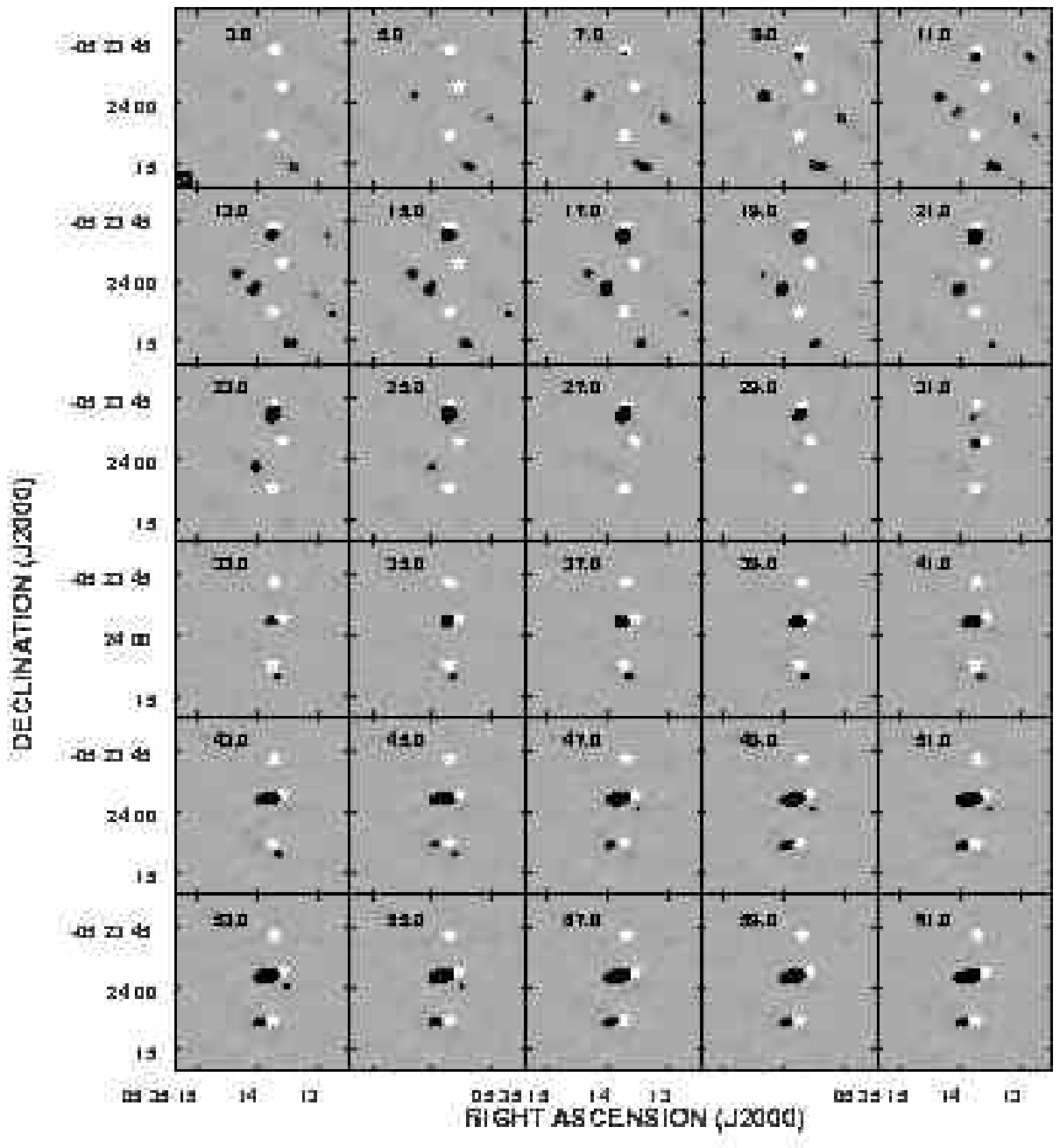} 
\caption{(Continued) Channel image of the SiO J=5$\rightarrow$4  redshifted emission of the 
molecular outflows. The emission is summed in velocity bins of 2 km s$^{-1}$. 
The central velocity is indicated in the upper right-hand corner of each panel 
(the systemic velocity of the ambient molecular cloud is about 5 km s$^{-1}$). 
The FWHM of the synthesized beam is shown in the lower left-hand corner of the first panel. 
The contours are 3, 4, 5, 6, 7, 8, 9, 10, 20, 30 $\times$ 0.07 Jy beam$^{-1}$ 
km s$^{-1}$, 
the rms noise of the image. The white stars indicates the position of the presumed exciting sources
of the SiO outflows. The upper white star denotes the position of the source 137-347, 
the middle star denotes the position of the source B or source 136-356, 
and the lower star denotes the position  of the source 137-408 (Gaume et al. 2005; 
Zapata et al. 2004b; Zapata et al. 2005).  }
\label{fig4}
\end{center}
\end{figure}

\clearpage


\begin{figure}
\begin{center}
\includegraphics[scale=1]{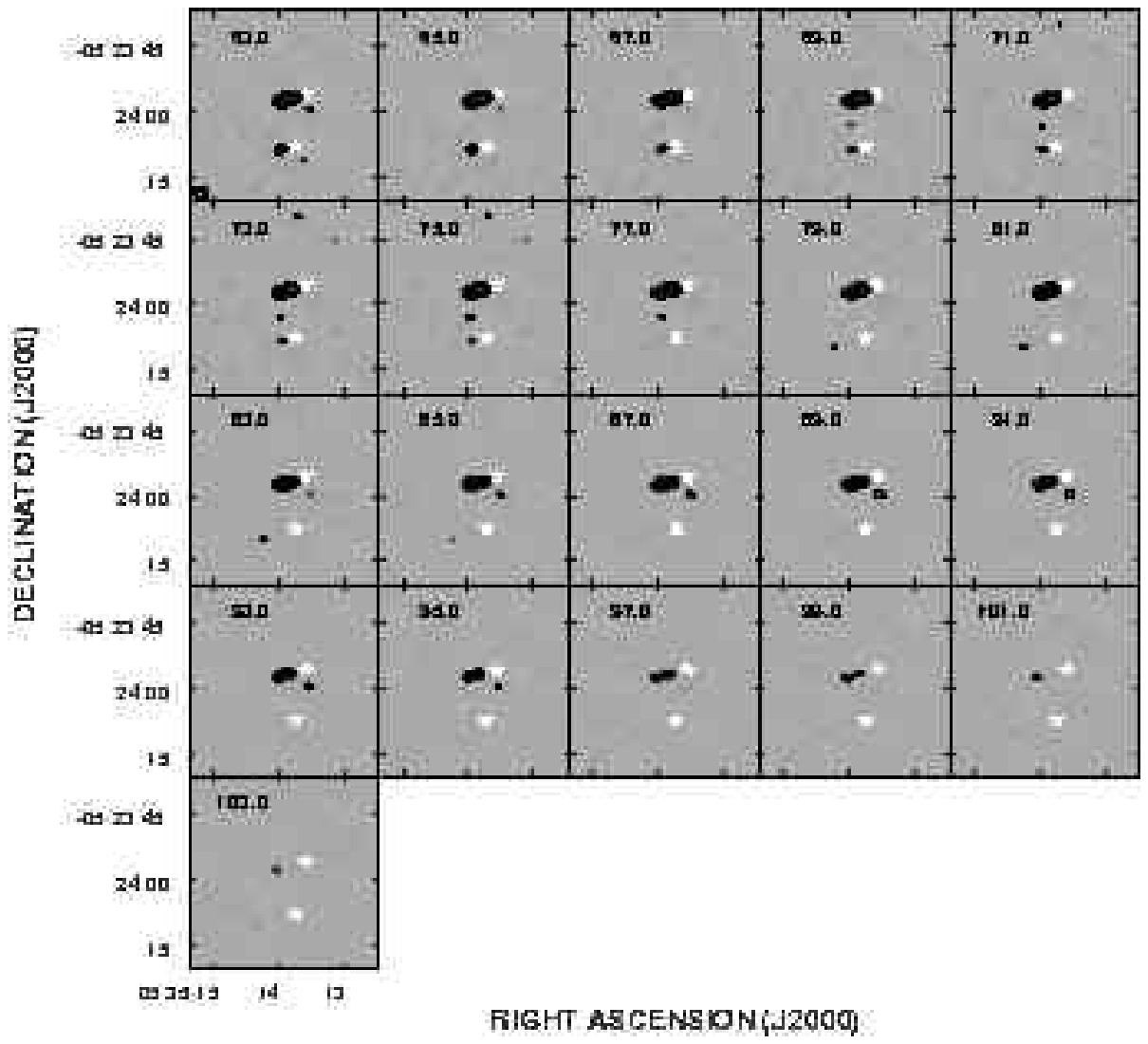} 
\caption{(Continued) Channel image of the SiO J=5$\rightarrow$4  redshifted emission of the 
molecular outflows. The emission is summed in velocity bins of 2 km s$^{-1}$. 
The central velocity is indicated in the upper right-hand corner of each panel 
(the systemic velocity of the ambient molecular cloud is about 5 km s$^{-1}$). 
The FWHM of the synthesized beam is shown in the lower left-hand corner of the first panel. 
The contours are 2, 3, 4, 5, 6, 7, 8, 9, 10, 20, and 30 $\times$ 0.07 Jy beam$^{-1}$
km s$^{-1}$, 
the rms noise of the image. The white stars indicate the position of the presumed exciting sources
of the SiO outflows.
 The upper white star denotes the position of the source B or source 136-356  
(Gaume et al. 1998; Zapata et al. 2005), while lower star denotes the position 
of the source 137-408 (Zapata et al. 2005). }
\label{fig5}
\end{center}
\end{figure}

\clearpage


\clearpage

\begin{figure}
\begin{center}
\includegraphics[scale=0.6, angle=0]{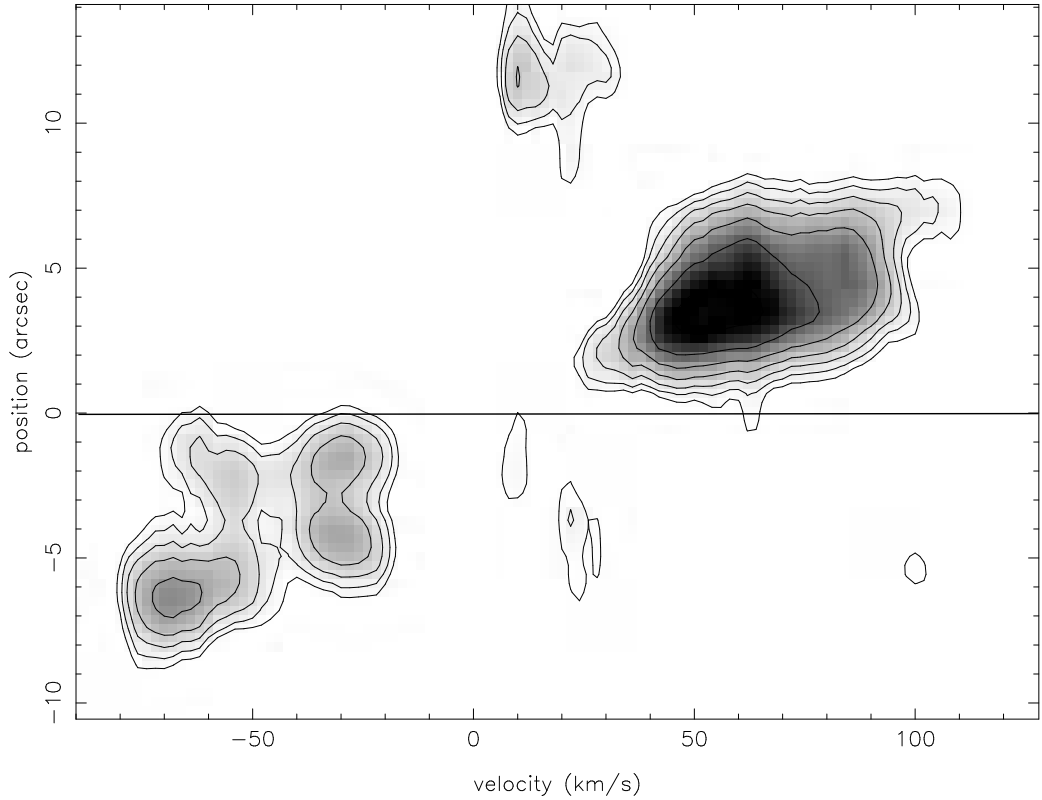} 
\caption{Position-velocity diagram of the SiO J=5$\rightarrow$4 thermal emission 
associated with the source 136-356 ,
computed along the jet axis (P.A. = 103$^{\circ}$). The data set was smoothed in velocity.
The lowest level corresponds to 0.08 Jy and the step is approximately logarithmic
(0.08, 0.13, 0.20, 0.30, 0.50, 0.80, and 1.2). 
The velocity and angular resolutions are 2 km s$^{-1}$ and $\sim$ 2$''$, respectively.
The systemic velocity of the ambient molecular cloud is about 5 km s$^{-1}$.
The horizontal dashed line indicates the position of the presumed exciting source.}
\label{fig9}
\end{center}
\end{figure}


\begin{figure}
\begin{center}
\includegraphics[scale=0.4, angle=0]{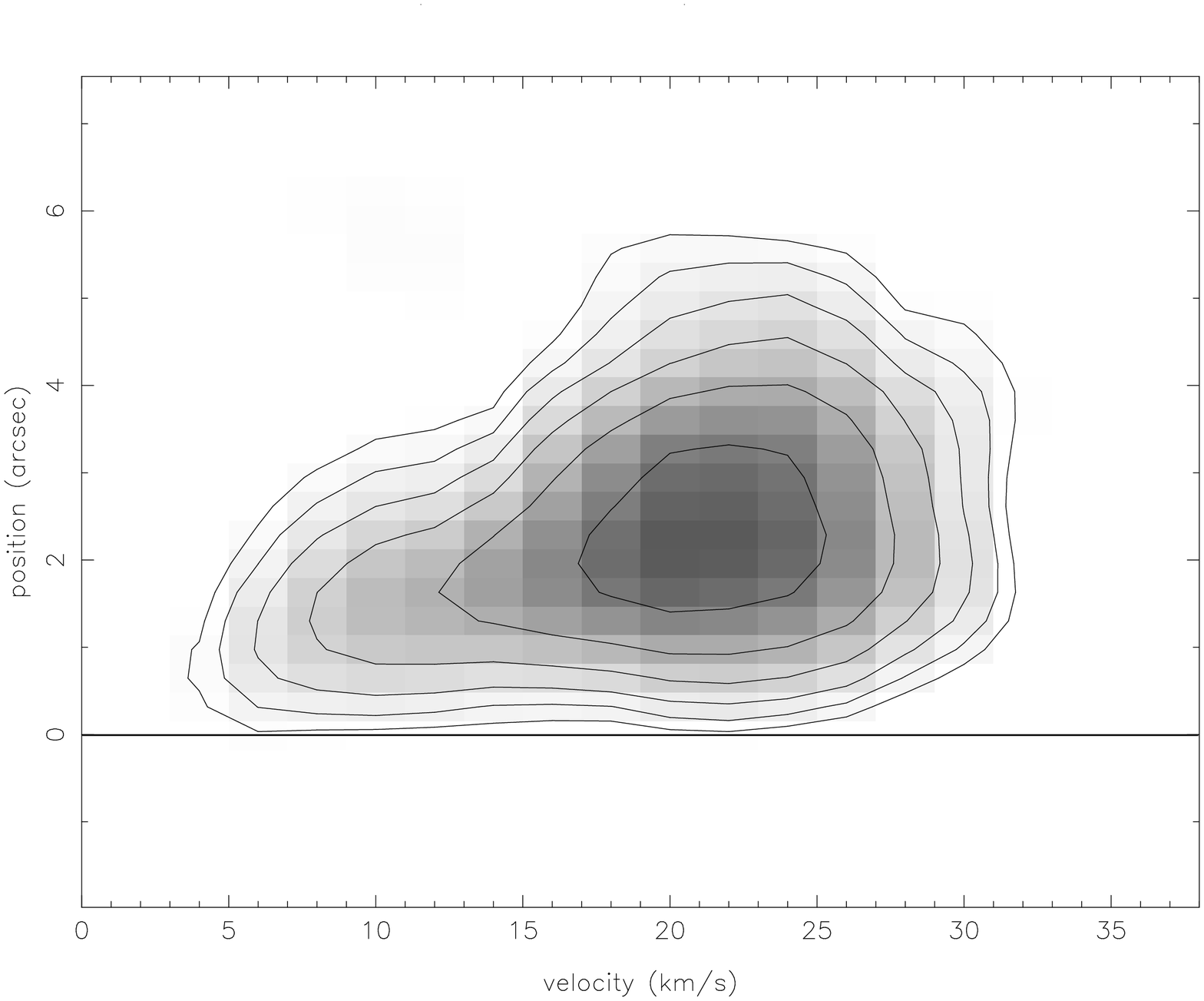} 
\caption{Position-velocity diagram of the SiO J=5$\rightarrow$4 thermal emission 
associated with the source 137-347,
computed along the jet axis (P.A. = 161$^{\circ}$). The data set was smoothed in velocity.
The lowest level corresponds to 0.08 Jy and the step is approximately logarithmic
(0.08, 0.13, 0.20, 0.30, 0.50, and 0.80). 
The velocity and angular resolutions are 2 km s$^{-1}$ and $\sim$ 2$''$, respectively.
The systemic velocity of the ambient molecular cloud is about 5 km s$^{-1}$.
The horizontal dashed line indicates the position of the presumed exciting source.}
\label{fig6}
\end{center}
\end{figure}

\clearpage

\begin{figure}
\begin{center}
\includegraphics[scale=0.6, angle=0]{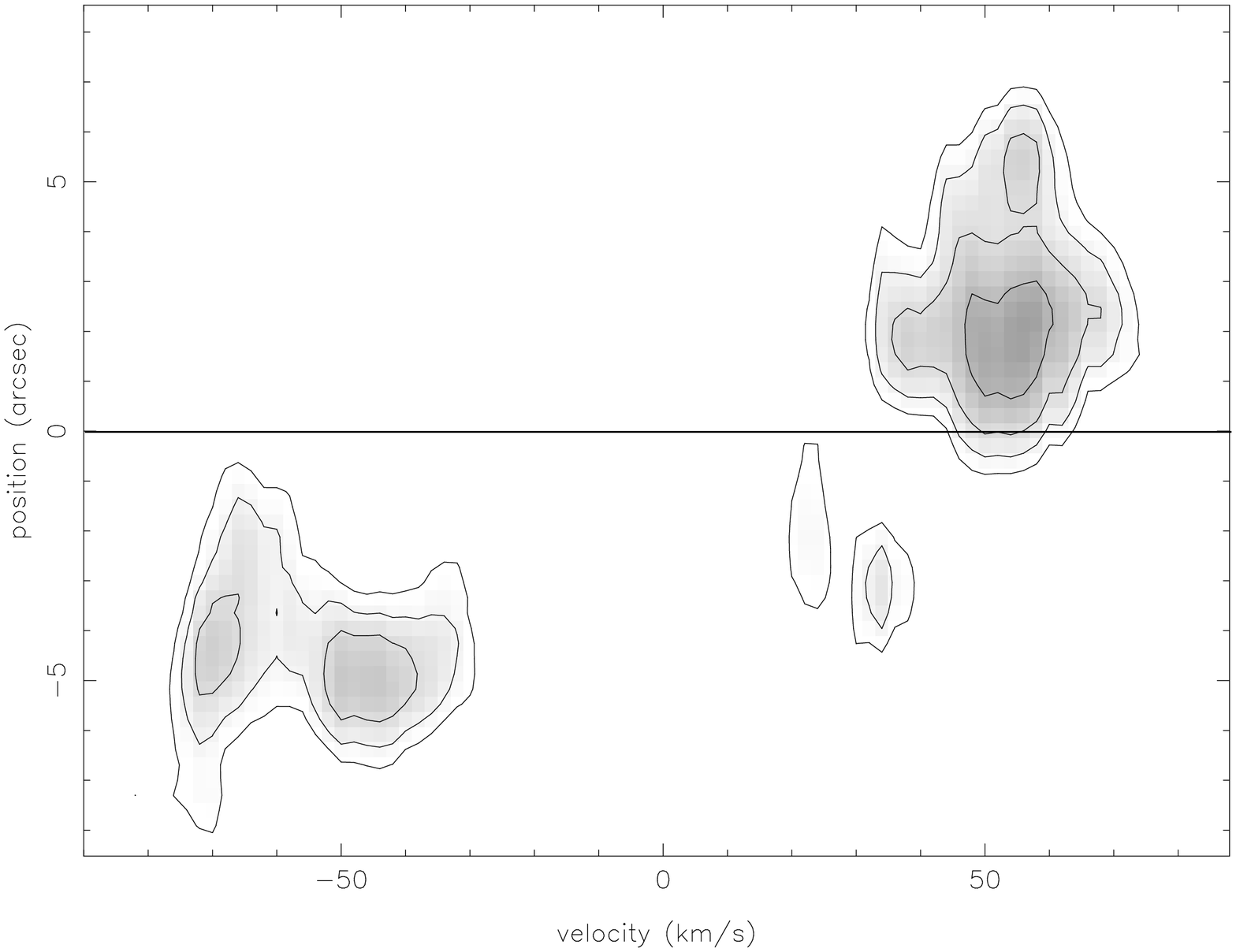} 
\caption{Position-velocity diagram of the SiO J=5$\rightarrow$4 thermal emission 
associated with the source 137-408,
computed along the jet axis (P.A. = 100$^{\circ}$). The data set was smoothed in velocity.
The lowest level corresponds to 0.07 Jy and the step is approximately logarithmic
(0.07, 0.11, 0.17, and 0.27). 
The velocity and angular resolutions are 2 km s$^{-1}$ and $\sim$ 2$''$, respectively.
The systemic velocity of the ambient molecular cloud is about 5 km s$^{-1}$.
The horizontal dashed line indicates the position of the presumed exciting source.}
\label{fig7}
\end{center}
\end{figure}

\clearpage

\begin{figure}
\begin{center}
\includegraphics[scale=0.6, angle=0]{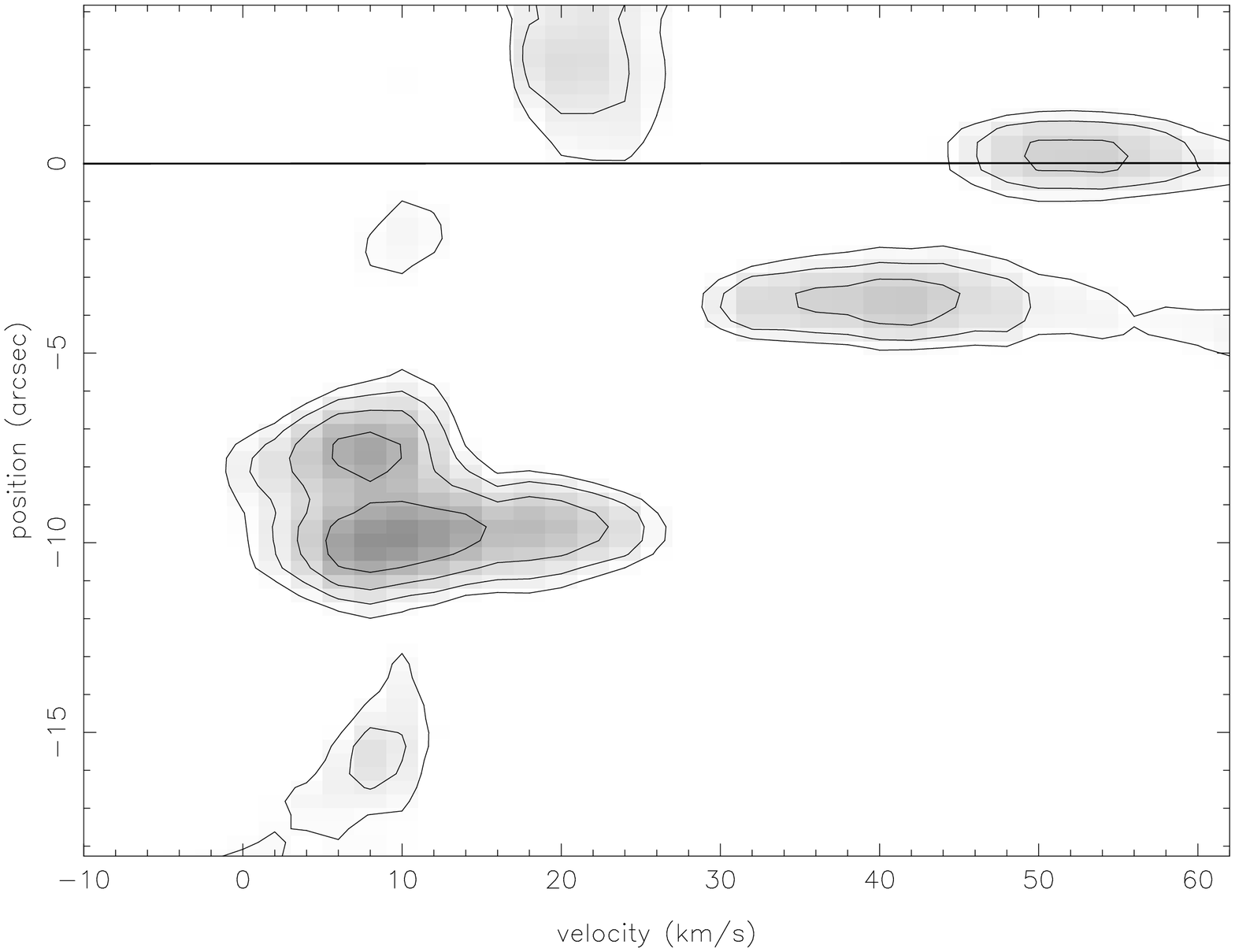} 
\caption{Position-velocity diagram of the SiO J=5$\rightarrow$4 thermal emission 
also associated with the source 137-408,
computed along the jet axis (P.A. = 34$^{\circ}$). The data set was smoothed in velocity.
The lowest level corresponds to 0.07 Jy and the step is approximately logarithmic
(0.07, 0.11, 0.17, and 0.27). 
The velocity and angular resolutions are 2 km s$^{-1}$ and $\sim$ 2$''$, respectively.
The systemic velocity of the ambient molecular cloud is about 5 km s$^{-1}$.
The horizontal dashed line indicates the position of the presumed exciting source.}
\label{fig8}
\end{center}
\end{figure}

\clearpage

\begin{figure}
\begin{center}
\includegraphics[scale=0.6, angle=0]{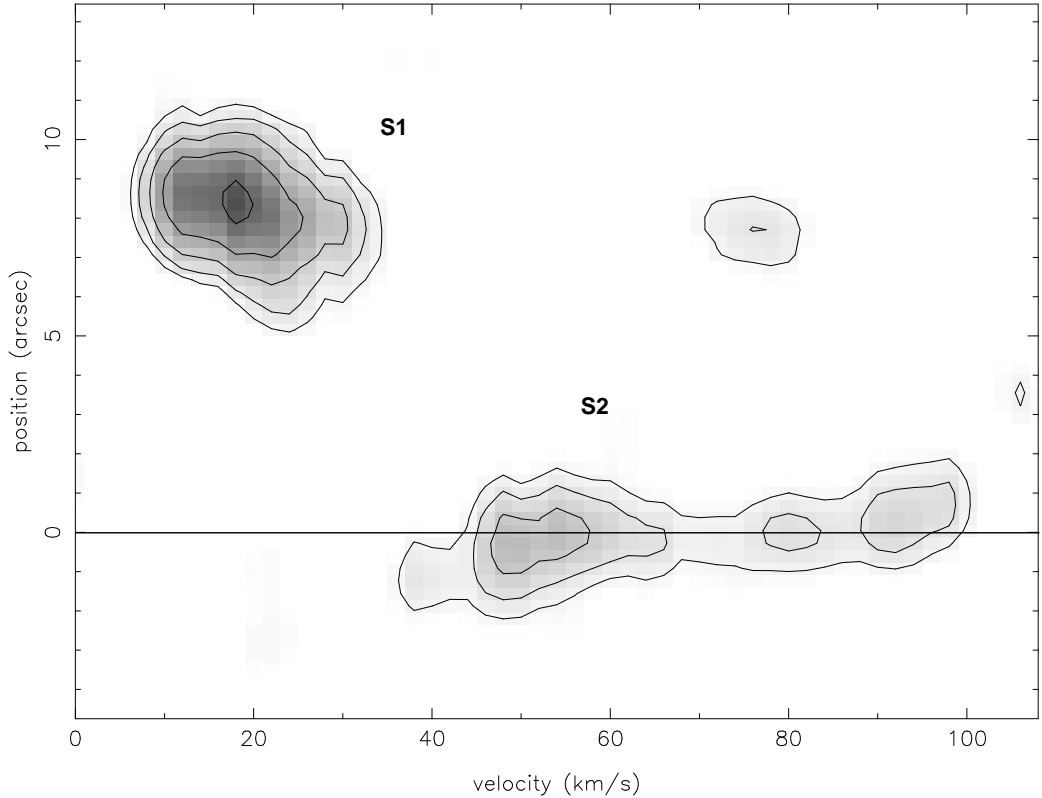} 
\caption{Position-velocity diagram of the SiO J=5$\rightarrow$4 thermal emission 
associated with the source 135-359.  It was computed with a P.A.=110$^{\circ}$. 
The data set was smoothed in velocity. The lowest level corresponds to 0.07 Jy and 
the step is approximately logarithmic (0.07, 0.11, 0.17, 0.27, and 0. 48). 
The velocity and angular resolutions are 2 km s$^{-1}$ and $\sim$ 2$''$, respectively.
The systemic velocity of the ambient molecular cloud is about 5 km s$^{-1}$.
The horizontal dashed line indicates the position of the presumed exciting source.}
\label{fig10}
\end{center}
\end{figure}

\end{document}